\documentclass[12pt]{article}
\usepackage{scicite}
\usepackage{times}
\usepackage{graphicx} 
\newenvironment{sciabstract}{%
\begin{quote} \bf}
{\end{quote}}

\newcounter{lastnote}
\newenvironment{scilastnote}{%
\setcounter{lastnote}{\value{enumiv}}%
\addtocounter{lastnote}{+1}%
\begin{list}%
{\arabic{lastnote}.}
{\setlength{\leftmargin}{.22in}}
{\setlength{\labelsep}{.5em}}}
{\end{list}}

\title{First Principles Free-Energy Theory of Solvation with Atomic Scale Liquid Structure}
 
\author
{Kendra Letchworth-Weaver,$^{1,2}$ Ravishankar Sundararaman,$^{1,3}$ T. A. Arias $^{1\ast}$\\
\\
\normalsize{$^{1}$Department of Physics, Cornell University, Ithaca, NY 14853}\\
\normalsize{$^{2}$Center for Nanoscale Materials, Argonne National Laboratory, Lemont, IL 60439}\\
\normalsize{$^{3}$Department of Materials Science and Engineering, Rensselaer Polytechnic Institute, Troy, NY 12180}\\
\\
\normalsize{$^\ast$To whom correspondence should be addressed; E-mail:  taa2@cornell.edu.}
}

\date{}

\begin{document}

\baselineskip24pt

\maketitle

\begin{sciabstract}
Quantum-chemical processes in liquid environments impact broad areas of science, from molecular biology to geology to electrochemistry. While density-functional theory (DFT) has enabled efficient quantum-mechanical calculations which profoundly impact understanding of atomic-scale phenomena, realistic description of the liquid remains a challenge. Here, we present an approach based on joint density-functional theory (JDFT) which addresses this challenge by leveraging the DFT approach not only for the quantum mechanics of the electrons in a solute, but also simultaneously for the statistical mechanics of the molecules in a surrounding equilibrium liquid solvent.  Specifically, we develop a new universal description for the interaction of electrons with an arbitrary liquid, providing the missing link to finally transform JDFT into a practical tool for the realistic description of chemical processes in solution.  This approach predicts accurate solvation free energies and surrounding atomic-scale liquid structure for molecules {\it and} surfaces in multiple solvents {\em without refitting}, all at a fraction of the computational cost of methods of comparable detail and accuracy. To demonstrate the potential impact of this method, we determine the structure of the solid/liquid interface, offering compelling agreement with more accurate (but much more computationally intensive) theories and with X-ray reflectivity measurements. 

\end{sciabstract}

\paragraph*{Introduction}
Quantum-mechanical processes in liquid environments are critical to a wide range of scientific disciplines and important, not only as the basis for several modern technologies, but also for sustaining life: redox processes at the electrochemical interface allow the operation of modern mobile devices based upon Li-ion batteries \cite{Abruna-pseudocap}. and chemical reactions catalyzed at the active sites of proteins are crucial to cell growth \cite{ProteinKinase}. First principles theory has great potential to advance science in these areas with new insights that can be obtained in no other way, but requires development of new methods to capture the relevant physical processes over multiple length and time scales. Specifically, capture of quantum-mechanical electron processes in a solute system simultaneously with the statistical-mechanical molecular arrangements in a solvent environment remains a challenge, particularly to do so sufficiently expeditiously to enable implementation of materials by design strategies \cite{MarzariNatureMaterials,RossmeislScience}. For example, in direct solar-to-hydrogen fuel conversion, understanding electron energy-level alignment between a semiconductor electrode and a water molecule undergoing photoelectrochemical splitting \cite{GratzelCell} requires determination of both the changes in the electronic wave function due to screening from the solvent environment and also the statistical average charge distribution in the solvent. Likewise, more rapid access to more reliable free energies and solvent configurations can enable improved identification of favorable configurations and reaction pathways for improved catalyst and enzyme design. We here introduce an approach to these challenges that provides a balance of robustness, accuracy, level of detail, and computational speed which complements existing techniques and accelerates the development of science impacted by solution chemistry.
 
\paragraph*{Previous work} 
Current approaches, at one extreme, include direct first-principles quantum-mechanical treatment of the solute and liquid environment \cite{CPMD,PayneCG}. The associated computational cost ranges from from extremely intensive to prohibitive because of the need to repeat many thousands of times quantum calculations of the entire system (solute plus a large set of representative solvent molecules) in order to statistically sample the relevant solvent configurations. Significant savings can be obtained by treating the liquid subsystem with a simpler theory, either using molecular dynamics with empirical potentials as in the QM/MM approach \cite{QMMM}, or using integral equation theories \cite{RISM1} within embedding theory \cite{CarterEmbedding} and the frozen density approximation \cite{Wesolowski93}. However, the former approach still requires statistical sampling, the latter does not address electronic and molecular motions in a unified framework, and both provide liquid structure, but calculation of free energies requires special care \cite{IntEqnSolEs}. At the opposite extreme, continuum solvation models replace the liquid response with that of a continuum dielectric \cite{PCM-Review,Truhlar}, albeit with significant empiricism, but provide little or no information about liquid structure.

To fill an important gap among the available approaches, we below build on a heretofore primarily formal development, for the first time enabling the theory to give simultaneously reliable free-energies and atomic-level liquid structure with costs comparable to the least expensive of the above approaches.  To bypass the prohibitive sampling of first principles quantum calculations while providing direct access to free energies and liquid structure, we employ the framework of Joint Density Functional Theory (JDFT) \cite{JDFT}. JDFT describes the solvent through statistically-averaged liquid structure and sets up an in-principle exact variational theory for the total \textit{free} energy $A$ in the form
\begin{equation}
A=\min_{n,\{N_\alpha\}} A[n,\{N_\alpha\}] = \min_{n,\{N_\alpha\}}( F_{HK}[n]+\Omega_{lq}[{N_\alpha}]+\Delta A[n,\{N_\alpha\}] ),
\label{eqn:mainDFT}
\end{equation}      
where  $n(r)$  is the quantum-statistical average electron density of the solute \textit{alone} and the $\{N_\alpha(r)\}$ are the statistically averaged densities of each atomic site $\alpha$ in the liquid (for pure water, $\alpha$=O,H). This free energy functional combines a quantum-mechanical density-functional theory (DFT) for the solute system $F_{HK}[n]$ \cite{KohnNobelPrize} with an atomically detailed classical DFT for the surrounding liquid environment $\Omega_{lq}[\{N_\alpha\}]$ \cite{PolCDFT}, using an exact, formally defined coupling functional $\Delta A[n,\{N_\alpha\}]$. Minimization of (\ref{eqn:mainDFT}) with respect to both electronic $n(r)$ and liquid $\{N_\alpha(r)\}$ degrees of freedom offers direct access to the temperature-dependent equilibrium free energy of the joint system as well as equilibrium densities of both components $n(r)$ and $\{N_\alpha(r)\}$, as proved in \cite{JDFT}, extending analogous theorems in \cite{Mermin-DFT,Capitani-UDFT}.

Akin to electronic DFT, the explicit form of the exact JDFT functional is unknown. The key ingredient missing in all previous work is an accurate yet efficient approximation for the coupling functional $\Delta A$. For the solute energy functional $F_{HK}[n]$, many established, efficacious density-functionals \cite{Mattson-exCorr} or quantum-chemical methods \cite{katie-QMC} are available. For the remaining two (liquid and coupling) functionals in (\ref{eqn:mainDFT}), one approach combines these into a single approximation for the liquid response, yielding a hierarchy of continuum solvation models \cite{PCM-Sahak,KLW2012}. Such JDFT-based approximations have computed successfully the solvation free energies of molecules and ions \cite{SGA13,SaLSA}, but, as with conventional solvation models, they ignore atomic-scale liquid structure. The one prior attempt to include explicit aqueous liquid structure in $\Delta A$ and $\Omega_{lq}$ was severely limited by overly simplistic liquid and coupling functionals incapable of capturing dielectric saturation, hydrogen bonding, and dispersion interactions\cite{JDFT}. Realistic, atomically-detailed classical DFT functionals $\Omega_{lq}$ that accurately capture effects pertinent to solvation (phase diagram, surface tension, nonlinear dielectric response) for multiple liquids have recently become available \cite{PolCDFT}, leaving a viable approximation for the coupling functional $\Delta A$ as the only barrier to practical application of a JDFT that simultaneously provides free-energies and detailed liquid structure.

\paragraph*{New Approach}
To develop this new, first principles theory of solvation, we introduce a universal, albeit approximate, JDFT coupling functional $\Delta A[n,\{N_\alpha\}]$ applicable to {\it any} quantum-mechanical solute in {\it any} solvent. The fundamental physical contributions (beyond mean-field electrostatic interactions) which $\Delta A$ must capture include mainly solute-solvent effects on the Coulomb, kinetic, and exchange-correlation energies of electrons. We thus construct the coupling functional as
\begin{equation}
\Delta A[n,\{N_\alpha\}]= \int \int \frac{\rho(r) \rho_{lq}(r')}{|r-r'|}\,d^3r d^3r' + \Delta \mathcal G[n,\{N_\alpha\}]
\label{eqn:CouplingFull}
\end{equation}
from the mean-field electrostatic interaction between solute and solvent charge densities $\rho(r)$ and $\rho_{lq}(r)$, and an additional term $\Delta \mathcal G [n,\{N_\alpha\}]$ capturing quantum-mechanical electronic effects. Following the embedding approach \cite{Hodak1,CarterEmbedding}, we compute $\Delta \mathcal G$ as the difference between the kinetic and exchange-correlation energies of the combined and isolated subsystems, as represented in standard electronic DFT as an electronic-density-only functional,
\begin{equation}
\Delta \mathcal{G}[n,\{N_\alpha\}]=\mathcal{G}[n+n_{lq}(\{N_\alpha\})]-\mathcal{G}[n]-\mathcal{G}[n_{lq}(\{N_\alpha\})],                                                           \label{eqn:CouplingEmbedding}
\end{equation}
where $n$ and $n_{lq}$ are the solute and solvent electron densities, respectively. The particular choice of $\mathcal{G}$ and reconstruction of the solvent electron density $n_{lq}(r)$ from the liquid structure $\{N_\alpha(r)\}$ determine the accuracy and computational feasibility of the new approach.  

We first consider a prescription for determining the electron density $n_{lq}(r)$ of the liquid directly from the atomic site densities $\{N_\alpha(r)\}$.  For this reconstruction, we decompose the first-principles computed electron density of a single liquid molecule, as screened by a solvent environment, into a sum of site-centered spherical atomic densities $n_\alpha(r)$  \cite{PolCDFT,Hodak1}. Summing the electron density contributions from this decomposition for each atomic site in the liquid, the electron density is computed as a convolution 
\begin{equation}
n_{lq}(r)=\sum_\alpha \int d^3R\hspace{4pt}N_\alpha(R) n_\alpha(r-R),                     \label{eqn:ConvCoupling}
\end{equation}
 as in \cite{Kaminski10}.
When performing the decomposition of solvent and solute electrons described above, it is essential that any electrons participating in charge transfer
reactions or covalent bonds be treated quantum-mechanically within $F_{HK}[n]$.

The remaining ingredient for the coupling functional (\ref{eqn:CouplingFull}) is to specify the practical approximation for $\mathcal{G}$ in (\ref{eqn:CouplingEmbedding}). Beginning with the simple local-density approximation (LDA) for both the kinetic and exchange-correlation contributions to $\mathcal{G}$, we equate the corresponding energy densities at each point in space to those of a uniform electron gas of equal density \cite{KohnNobelPrize}. Additionally, including long-range correlation effects (such as dispersion) in $\mathcal{G}$ can be critical to accurate calculation of free energies and liquid structure (especially for hydrophobic or non-polar solutes such as solvated graphene or alkane molecules). We incorporate dispersion interactions through non-local pair-potential van der Waals corrections between the atoms of the solute and the atoms of the liquid, following \cite{Grimme} but generalizing to a continuum description by integrating over the atomic site densities of the liquid $\{N_\alpha\}$\cite{SaLSA}. In contrast to previous JDFT work, this approach employs a self-consistent liquid density with accurate atomic-scale structuring, rather than a near-featureless continuum. As per \cite{Grimme}, this correction includes a single adjustable prefactor, which we set to $s_6=0.488$ based on the resulting molecular solvation energies \cite{SGA13}.  Future coupling functionals could entirely eliminate this adjustable parameter by using accurate, albeit more expensive, orbital-free van der Waals density-functionals \cite{Soler-vdW}. 

In summary, self-consistent minimization of the electronic energy of a solute coupled to an atomically detailed liquid through the above orbital-free density-functional theory is a novel milestone. Furthermore, this method is now implemented in the fully featured JDFTx software \cite{JDFTx-SoftwareX}, enabling first principles solvation calculations in a planewave basis set for both isolated (molecules) and extended (surfaces) systems. These developments bring the full promise of JDFT to fruition and make the theory practical for a wide range of applications. 

\begin{figure}
\centering
\includegraphics[width=\linewidth]{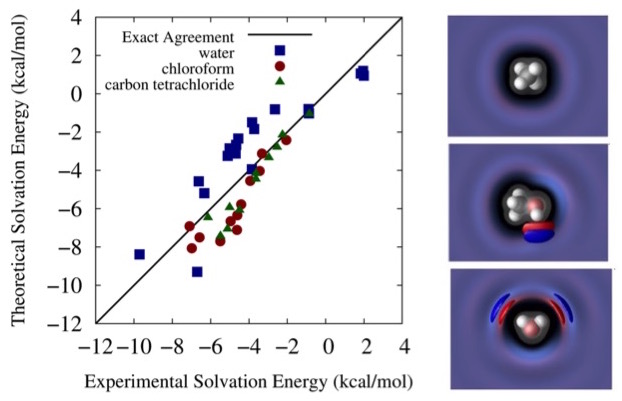}
\caption{(left) JDFT-predicted free energies of solvation compared to experimental measurements \cite{Truhlar}. (right) liquid water structure (O density in red, H density in blue) around quantum-mechanical  solutes of increasing dipole moment (top: methane, center: methanol, bottom: water)}
\label{fig:SolEs}
\end{figure}

\paragraph*{Results}
To verify that our coupling functional captures the subtle interplay among electrostatics, cavity formation, and polarizability in disparate solvents, we consider free energies of solvation for eighteen small, organic molecules in water (H$_2$O), twelve in chloroform (CHCl$_3$), and eleven in carbon tetrachloride (CCl$_4$). We choose these solvents to represent a sequence, from water, with a large dipole moment and electrostatic effects dominating the solvation energies, through (fully nonpolar) carbon tetrachloride for which cavity formation and polarizability dominate.  Despite the significant differences among these solvents, a single, global choice of \textit{a priori} coupling functional predicts solvation free energies (Figure \ref{fig:SolEs}) within a mean absolute error of only 1.18 kcal/mol from experimental measurements\cite{Truhlar} (rms error 1.42 kcal/mol). Such near-chemical accuracy, with errors approaching the scale of room-temperature thermal fluctuations, indicates that JDFT successfully captures the underlying physics for a variety of solute and solvent types. Indeed, we predict the near constant solvation energy amongst the alkanes (methane, ethane, and propane) in water, a challenging task involving a careful balance of the increasing cavity formation energy with the addition of more carbon atoms against the decreasing energy associated with increasing dispersion interactions.

\begin{figure}
\centering
\includegraphics[width=\linewidth]{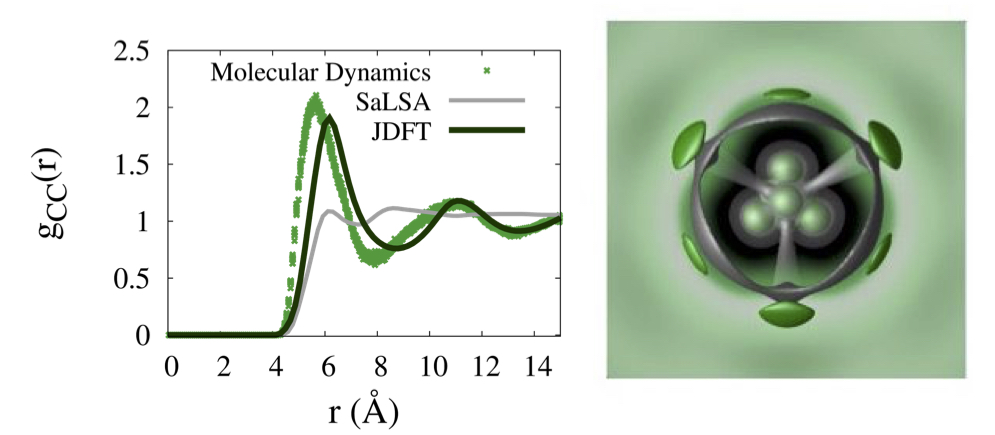}
\caption{(left) Carbon-carbon radial distribution function in bulk liquid CCl$_4$ from molecular dynamics \cite{PolCDFT} (green points), JDFT (black curve), and a nonlocal polarizable continuum model \cite{SaLSA} (grey curve). (right) Three-dimensional structure of the self-solvation of CCl$_4$ (C density in silver, Cl density in green, electron density as grey clouds). }
\label{fig:CCl4_structure}
\end{figure}

JDFT offers not only excellent performance for free energies but also, through the variational principle, direct access to equilibrium densities of each atomic species in the liquid environment. Figure \ref{fig:SolEs}(right) shows the JDFT-predicted, atomically-detailed solvation shell structures in liquid water, with significant organization around the polar solute molecules methanol and water (self-solvation), as compared to the nonpolar solute methane. The JDFT approach accurately represents the position and number of molecules in the first solvation shell for the challenging system of water in water, though the secondary solvation structure remains challenging due to the present lack of liquid functionals which capture subtle hydrogen bonding effects. (See SI for details.) However, JDFT does capture both primary and secondary solvation structure accurately in nonaqueous liquid: Figure \ref{fig:CCl4_structure} directly compares the carbon-carbon radial distribution function in carbon tetrachloride as obtained from JDFT and from classical molecular dynamics with thermodynamic sampling \cite{PolCDFT}.  In contrast, state-of-the-art continuum solvation theories (such as the nonlocal SaLSA model \cite{SaLSA}) offer little to no liquid structure. 

\begin{figure}
\centering
\includegraphics[width=\linewidth]{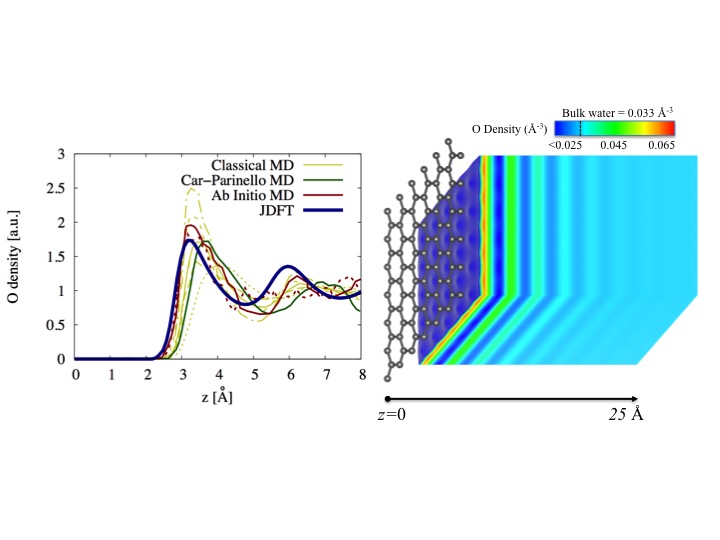}
\caption{(left) Comparison of aqueous liquid structure near graphene as predicted by JDFT, and classical and {\it ab initio} molecular dynamics \cite{Galli-graphene,Gogotsi-graphene,Chandra-graphene}. (right) Liquid water structure near graphene sheet (carbon atoms grey, oxygen site density as color map).}
\label{fig:graphene_structure}
\end{figure}

Furthermore, JDFT provides reliable predictions of liquid structure for extended systems, making it an efficient tool to study the fundamental science underlying the solid-liquid interface, such as in electrochemical devices. Figure \ref{fig:graphene_structure} compares the density of oxygen sites in water as a function of perpendicular distance $z$ to a graphene sheet as predicted with JDFT and classical and {\it ab initio} molecular dynamics (MD) \cite{Galli-graphene,Gogotsi-graphene,Chandra-graphene}. The assorted classical MD predictions depend strongly upon the modeled oxygen-carbon interaction strength, resulting in wide variations in the height and location of the first peak.  {\it Ab initio} MD (AIMD) predictions of the first peak are remarkably consistent, but second peak predictions vary substantially due to limitations on numbers of atoms and sampling times imposed by the significant computational costs of these heroic calculations. In contrast, \textit{within a few minutes on a desktop workstation}, JDFT predicts the temporally and spatially averaged liquid structure next to graphene with accuracy comparable to the discrepancy between the published AIMD results.

\begin{figure}
\centering
\includegraphics[width=\linewidth]{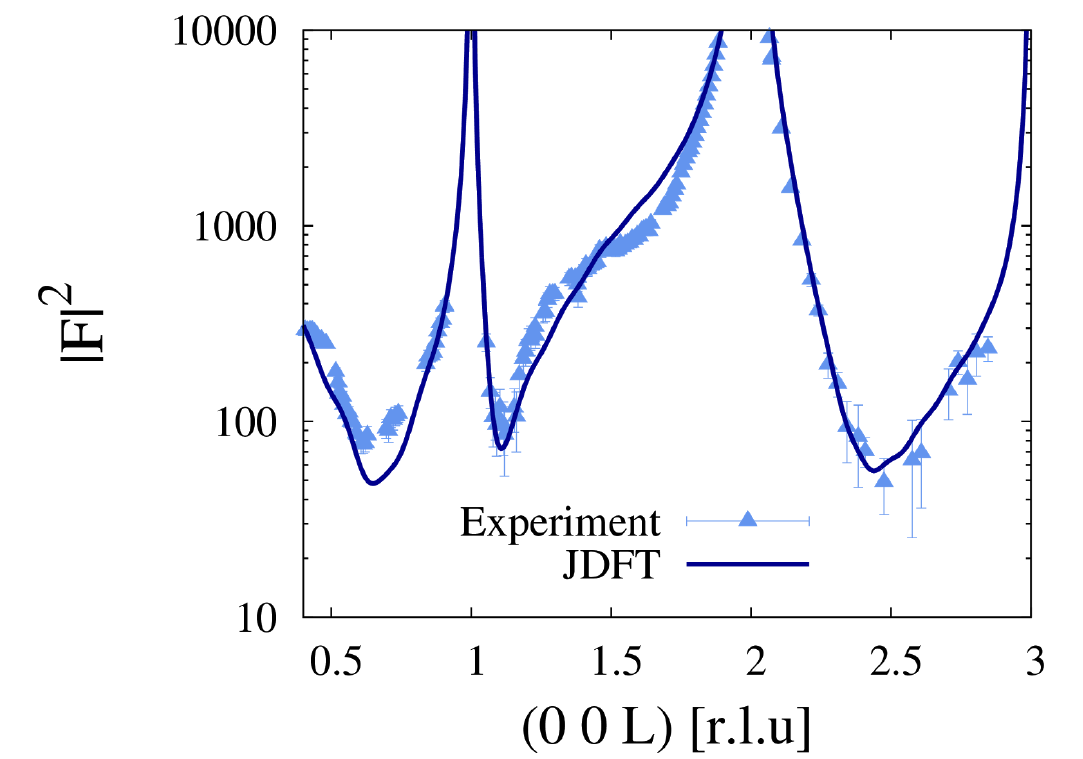}
\caption{X-ray reflectivity as a function of specular momentum transfer measured by experiment (triangles) \cite{STO_JACS} and computed from JDFT with the solvent-independent coupling functional (solid line).}
\label{fig:STO_CTR}
\end{figure}

Finally, our predicted liquid structures near semiconducting oxide surfaces are in excellent agreement with experiments. Specifically, from the resulting JDFT interfacial structure, we can predict the X-ray reflectivity of SrTiO$_3$ (001) in aqueous solution and compare directly to the corresponding experiments \cite{STO_JACS}. (The previous work on this interface\cite{STO_JACS} lacked the polarizability of the water molecules in the liquid functional \cite{PolCDFT} and the van der Waals corrections in the coupling functional (\ref{eqn:CouplingFull}) and thus could not predict accurate free-energies nor liquid structure near the surface within the single global framework presented herein.)  Figure \ref{fig:STO_CTR} shows remarkable agreement between theory and experiment without any adjustment of parameters to the X-ray signal, establishing that the JDFT approach, with a single global coupling functional, not only provides accurate free energies, but also liquid structure at solvated solid surfaces.  

\paragraph*{Conclusion}
We here provide the missing link to transform a heretofore purely formal, theoretical development into a practical, efficient tool to study quantum systems in contact with liquid environments. Specifically, we provide the first joint density-functional theory (JDFT) capable of predicting free energies and atomically detailed liquid structure simultaneously, within a single, global framework for multiple solvents, all at a fraction of the computational cost of methods of comparable detail and accuracy. The excellent performance of the JDFT approach compared with both experiment and state-of-the-art theory suggests that it has matured into a predictive solvation theory which can bridge the gap between computational efficiency and accuracy.  We anticipate JDFT to impact broad areas of Science involving solution chemistry in much the same way that traditional density-functional theory has impacted solid-state chemistry and materials science.
	
\bibliography{references,referencesThesis}

\bibliographystyle{Science}

\begin{scilastnote}
\item 
This material is based, in part, on work supported by the Energy Materials Center at Cornell, an Energy Frontier Research Center funded by the U.S. Department of Energy, Office of Science, Office of Basic Energy Science under Award Number DE-SC0001086. 

Use of the Center for Nanoscale Materials, an Office of Science user facility, was supported by the U. S. Department of Energy, Office of Science, Office of Basic Energy Sciences, under Contract No. DE-AC02-06CH11357.

\end{scilastnote}

\clearpage

\end{document}